\documentclass[prb,aps,twocolumn,showpacs,nofootinbib]{revtex4}

\usepackage{graphicx}
\usepackage{amsmath}
\usepackage{wasysym}
\usepackage{amsfonts}
\usepackage{amssymb}
\usepackage{dcolumn} 
\usepackage{bm} 
\usepackage{color}

\usepackage{latexsym}

\begin{document}

\title{ Spin Superfluidity in Coplanar Multiferroics}

\author{Wei Chen$^{1}$ and Manfred Sigrist$^{2}$} 

\affiliation{$^{1}$Max-Planck-Institut f$\ddot{u}$r Festk$\ddot{o}$rperforschung, Heisenbergstrasse 1, D-70569 Stuttgart, Germany 
\\
$^{2}$Theoretische Physik, ETH-Z\"urich, CH-8093 Z\"urich, Switzerland}

\date{\rm\today}

\begin{abstract}

{ Multiferroics with coplanar magnetic order are discussed in terms of a superfluid condensate, with special emphasis on spin supercurrents created by phase gradients of the condensate and the effect of external electric fields. By drawing the analogy to a superconducting condensate, phenomena such as persistent currents in rings, the Little-Parks effect, fluxoid quantization, the Josephson-like effect through spin domain walls, and interference behavior in a SQUID-like geometry are analyzed for coplanar multiferroics.  } 

\end{abstract}






\maketitle

\section{Introduction} 

Spintronics with its potential to substitute usual electronic devices is a fast developing and promising field of condensed matter physics. Important discoveries such as spin pumping \cite{Tserkovnyak02}, spin-transfer torque \cite{Slonczewski96,Berger96}, (inverse) spin Hall effect \cite{Dyakonov71,Hirsch99,Saitoh06,Valenzuela06,Kimura07}, and various other phenomena make the generation and detection of dissipative spin currents a reality. More recently, the notion of spin superfluidity has emerged as another useful concept. Certain magnetically ordered phases can be represented as superfluid condensates\cite{Halperin69,Chandra90,Sonin10,Takei13}, which support dissipationless spin supercurrents when the magnetic order involves a certain texture. Moreover, the generic form of canonical momentum associated with the Aharonov-Casher (AC) effect \cite{Aharonov84} or Rashba spin-orbit coupling (SOC), implies that the cross product ${\boldsymbol\mu}\times{\bf E}$ between magnetic moment of the spin supercurrent and an applied electric field plays a role analogous to a gauge field. Despite this analogy with the vector potential, there is no freedom of gauge here, since both ingredients of the cross product are directly measurable quantities\cite{Chen13}. The corresponding condensate and gauge field formalism provides the possibility of controlling the spin superfluidity by electric fields. This is true, in particular, if there is a strong coupling of electric polarization and magnetic degrees of freedom, as found in multiferroics. 

In this article, we demonstrate that in coplanar multiferroic insulators strong magnetoelectric coupling, incorporated in ${\boldsymbol\mu}\times{\bf E}$, can modify the phase responsible for spin supercurrents \cite{Hikihara08,Okunishi08,Kolezhuk09}. Using for the description of the coplanar magnetic order a complex condensate wave function \cite{Chandra90,Sonin10}, one can show that the Dzyaloshinskii-Moriya (DM) interaction controlled by an electric field \cite{Shiratori80} introduces a behavior like a gauge field for a superconductor. Moreover, in a coplanar magnet the existence of spin supercurrents follows from the commutation relation \cite{Villain74} $\left[\theta,S^{y}\right]=i$, in an analogous way to the phase-density commutation relation $\left[\varphi,{\hat N}\right]=i$ in superconductors, where $\theta$ is the planar angle of the orientation of the ordered spins confined in the $xz$-plane. Based on this analogy, we show that several concepts known for superconductors can be simply transferred into their spintronic analogues. 

This article is organized as follows. In Sec.~II, we give a detailed discussion of the mapping of a coplanar magnet to a complex condensate, and explain how the electric field induces a behavior analogous to the $U(1)$ gauge field in superconductors. In Sec.~III, various phenomena are analyzed for coplanar multiferroics by drawing analogy with superconductors, including persistent currents in Corbino rings, the Little-Parks effect, fluxoid quantization, the Josephson-like effect through spin domain walls, and quantum interference of spin supercurrents. Furthermore, the possible experimental realization in known materials is briefly discussed. Sec.~IV summarizes our calculations, and points at possible applications as well as the difference between multiferroics and superconductors.

\section{Superfluid description of coplanar magnetic order}

We consider a coplanar multiferroic system including DM interactions described by the following Hamiltonian on a square lattice,
\begin{eqnarray}
H=\sum_{i}\sum_{\alpha=a,c}J{\bf S}_{i}\cdot{\bf S}_{i+\alpha}-{\bf D}_{\alpha}\cdot\left({\bf S}_{i}\times{\bf S}_{i+\alpha}\right)
\label{DM_Hamiltonian}
\end{eqnarray}
where ${\boldsymbol\alpha}=\left\{{\bf a},{\bf c}\right\}$ is the unit vector of the square lattice, and we restrict the discussion to ${\bf D}_{\alpha}\parallel{\bf\hat{y}}$, perpendicular to the plane, that can vary locally. A uniform ${\bf D}_{\alpha}$ causes coplanar spiral order. Other spin configurations, such as vortices, can be induced by locally varying ${\bf D}_{\alpha}$ \cite{Mostovoy06}. The DM interaction on the $\alpha$-bond ${\bf D}_{\alpha}=w{\bf E}\times{\boldsymbol\alpha}$ can be viewed as being induced by an ${\bf E}$ field either intrinsic to the material and/or applied externally \cite{Shiratori80}. The classical energy of a ferromagnetic spiral $J<0$ of local wave vector ${\bf Q}$ is 
\begin{eqnarray}
E=\sum_{i}\sum_{\alpha}\tilde{J}_{\alpha}S^{2}\cos\left({\bf Q}\cdot{\boldsymbol\alpha}-\theta_{E\alpha}\right)
\end{eqnarray}
where $\tilde{J}_{\alpha}=-\sqrt{J^{2}+D_{\alpha}^{2}}$, and $\theta_{E\alpha}=\sin^{-1}\left(D_{\alpha}/|\tilde{J}_{\alpha}|\right)$. 
For an antiferromagnetic spiral $J>0$, upon rotation in one sublattice ${\bf S}_{i}\rightarrow(-1)^{i}{\bf S}_{i}$, the classical energy $E$ and the planar coupling $\tilde{J}_{\alpha}$ are the same as for the ferromagnetic spiral, but with $\theta_{E\alpha}=-\sin^{-1}\left(D_{\alpha}/|\tilde{J}_{\alpha}|\right)$. Including the contribution from electric polarization ${\bf P}=\epsilon_{0}\chi_{e}{\bf E}$, the free energy density for small deviations ${\bf Q}\cdot{\boldsymbol\alpha}\approx\theta_{E\alpha}$ is given by\begin{eqnarray}
f&=&f_{0}+\frac{|\tilde{J}_{a}|+|\tilde{J}_{c}|}{v}\left(\frac{\tau}{\tau_{c}}-1\right)S^{2}+\frac{\beta}{2v^{2}}S^{4}
\nonumber \\
&+&\sum_{\alpha=a,c}\frac{|\tilde{J}_{\alpha}|S^{2}}{2v}\left({\bf Q}\cdot{\boldsymbol\alpha}-\theta_{E\alpha}\right)^{2}+\frac{|{\bf P}|^{2}}{2\epsilon_{0}\chi_{e}}
\label{GL_DM}
\end{eqnarray}
where $f_{0}$ is a constant and $\beta >0 $ is the phenomenological coefficient of the $S^{4}$ term \cite{Kittel80}. We assume that the coplanar order is stabilized at finite temperatures $\tau$ below $ \tau_c $ because of weak coupling between planes.

We use the mapping
\begin{eqnarray}
\langle{\bf S}\rangle=S\left(\sin\theta,0,\cos\theta\right)=\sqrt{v}\left(\Im\psi,0,\Re\psi\right)
\label{the_mapping}
\end{eqnarray}
where $v$ is the volume of the 3D unit cell \cite{Chandra90}. The phase of the condensate wave function $\psi=|\psi|e^{i\theta}$ is the planar angle of the spins. In the coherent state formalism for spin $1/2$, this corresponds to $|s\rangle=\left(\cos\frac{\theta}{2},\sin\frac{\theta}{2}\right)^{T}$ and $\langle{\bf S}\rangle=\langle s|{\bf S}|s\rangle$. Within this description, Eq. (\ref{GL_DM}) assumes a form similar to the Ginzburg-Landau(GL) free energy of a conventional superconductor
\begin{eqnarray}
f&=&f_{0}+\left(|\tilde{J}_{a}|+|\tilde{J}_{c}|\right)\left(\frac{\tau}{\tau_{c}}-1\right)|\psi|^{2}+\frac{\beta}{2}|\psi|^{4}
\nonumber \\
&&+\frac{1}{2m}|\left(\frac{\hbar}{i}{\boldsymbol\nabla}-\lambda{\boldsymbol\mu}\times{\bf E}\right)\psi|^{2}+\frac{|{\bf P}|^{2}}{2\epsilon_{0}\chi_{e}}
\label{GL_spin_Josephson}
\end{eqnarray}
where $\lambda{ \boldsymbol\mu}\times{\bf E} $ corresponds to the magnetoelectric coupling of spin current and electric field, where ${\boldsymbol\mu} $ is
the magnetic moment of the spin supercurrent (${\boldsymbol\mu}\parallel{\bf{\hat y}}$). Assuming the magnitude $|\psi|$ remains constant ignoring quantum fluctuations of ${\bf S}$, the direct comparison with Eq. (\ref{GL_DM}) in the long-wavelength limit $\tilde{J}_{\alpha}\approx J$ yields $\hbar^{2}/m =|J||{\boldsymbol\alpha}|^{2}$, $w=\lambda|J||{\boldsymbol\mu}|/\hbar$ connecting the coupling constants  $\lambda $ with the DM interaction. 
The GL coherence length is $\xi=|{\boldsymbol\alpha}|\left(1-\tau/\tau_{c}\right)^{-1/2}/2$.

From Eq.~(\ref{DM_Hamiltonian}), the classical energy on the bond $\left\{i,i+a\right\}$ is $E=\tilde{J}S^{2}\cos\left(\delta\theta-\theta_{E}\right)$, where $\delta\theta\equiv\theta_{i+a}-\theta_{i}= {\bf a} \cdot {\boldsymbol\nabla}\theta$ and $\theta_{E}\equiv\lambda\left({\boldsymbol\mu}\times{\bf E}\right)\cdot {\bf a}/\hbar$. If $\delta\theta\neq\theta_{E}$, a spin supercurrent due to { virtual} hopping of electrons is created on the bond \cite{Katsura05,Bruno05}
\begin{eqnarray}
J_{s}=\langle \dot{S}_{i}^{y}\rangle=-\langle \dot{S}_{i+a}^{y}\rangle
=\tilde{J}S^{2}\sin\left(\delta\theta-\theta_{E}\right)=-\frac{1}{\hbar}\frac{\partial E}{\partial\delta\theta}\;,
\label{spin_current_from_Hamiltonian}
\end{eqnarray}
which transports out-of-plane angular momentum ${\boldsymbol{\hat \mu}}\parallel{\bf {\hat y}}$. Eq.~(\ref{spin_current_from_Hamiltonian}) follows the analogous relation as between energy and Josephson current, because commutation relations $\left[\varphi,\hat{N}\right]=i$ and $\left[\theta,S^{y}\right]=i$ yield the analogous formulation. For small deviations $\delta\theta\approx\theta_{E}$ in the continuous limit, the spin supercurrent can be described as a supercurrent of the condensate
\begin{eqnarray}
J_{s}={\bf J}_{s}\cdot{\bf {\hat a}}=-\frac{v|\psi|^{2}}{am}{\bf \hat{a}}\cdot\;\left(\hbar{\boldsymbol\nabla}\theta-\lambda{\boldsymbol\mu}\times{\bf E}\right)\;.
\label{spin_current_from_GL}
\end{eqnarray}
where the ${\boldsymbol\nabla}\theta$ term represents the vector chirality \cite{Hikihara08,Okunishi08,Kolezhuk09}. 
Empirically, the continuity equation for systems with strong SOC is written as 
\begin{eqnarray}
a{\boldsymbol\nabla}\cdot{\bf J}_{s}+\langle \dot{S}^{y}\rangle+\frac{\langle S^{y}\rangle}{\tau_{s}}=0\;.
\label{continuity_condition}
\end{eqnarray}
where $\tau_{s}$ is the spin relaxation time \cite{Shi06}. In the following, we restrict ourselves to the equilibrium spin supercurrent under the condition $\langle \dot{S}^{y}\rangle=0$ and $\langle S^{y}\rangle=0$, i.e. the spin order remains static and coplanar, such that ${\boldsymbol\nabla}\cdot{\bf J}_{s}=0$ and the spin supercurrent is a stable ground state property  as long as the system size is smaller than the spin relaxation length.

\begin{figure}[ht]
\begin{center}
\includegraphics[clip=true,width=0.6\columnwidth]{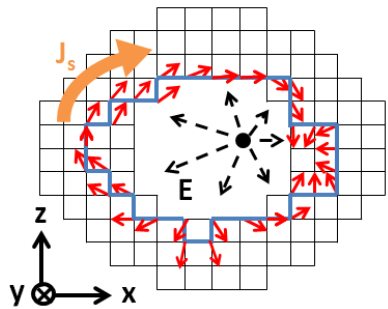}
\caption{ (color online) Corbino ring of a square lattice multiferroic. A line charge piercing the ring creates a planar electric field $ {\bf E} $ and a DM interaction vector $ {\bf D}_{\alpha} $ pointing out of the plane. A possible resulting planar spin spiral state is indicated by (red) arrows along a (blue) path around the ring (here winding number $ n=2 $). In this configuration a spin current ${\bf J}_{s}$ can circulate around the ring. }
\label{fig:2D_Corbino}
\end{center}
\end{figure}

\section{Phenomena analogous to superconductivity}

\subsection{Persistent spin current and fluxoid quantization}

Although ${\boldsymbol\mu}\times{\bf E}$ has no gauge freedom, it does cause effects analogous to those caused by the { $U(1)$ gauge} field in a superconductor. The first example is the persistent current in a ring geometry. For a system with open boundary conditions and a uniform electric field ${\bf E} $, a coplanar spiral ground state minimizing Eq. (\ref{GL_DM}) or (\ref{GL_spin_Josephson}) follows from $\hbar{\boldsymbol\nabla}\theta=\lambda{\boldsymbol\mu}\times{\bf E}$. Thus, no spin current occurs in this ground state. One way to create spin current is by imposing periodic boundary condition, such that the single-valuedness of the spin wave function enforces $\hbar{\boldsymbol\nabla}\theta\neq\lambda{\boldsymbol\mu}\times{\bf E}$ in Eq. (\ref{spin_current_from_GL}). This is very much in the same way a persistent supercurrent flows by a winding of the phase around a superconducting ring threaded by a magnetic field\cite{Byers61,Buttiker83}. To illustrate this, we consider a Corbino ring extracted from a 2D square lattice { multiferroics}, as shown in Fig. \ref{fig:2D_Corbino}. A uniformly charged line piercing the ring produces a radial, inhomogeneous ${\bf E}$ field. The single-valuedness of the spin wave function requires
\begin{eqnarray}
\oint{\boldsymbol\nabla}\theta\cdot d{\bf l}=2\pi n
\label{single_valuedness}
\end{eqnarray}
on any loop that encloses the hole. Here the integer $n$ is the phase winding number on the closed path around the ring. Eqs. (\ref{spin_current_from_GL}) and (\ref{single_valuedness}) yield a quantization condition similar to the fluxoid quantization in a superconducting ring
\begin{eqnarray}
{\tilde\Phi}_{E}={\boldsymbol{\hat\mu}}\cdot{\bf\Phi}_{E}+\frac{\hbar^{2}}{aJS^{2}|{\boldsymbol\mu}|\lambda}\oint{\bf J}_{s}\cdot d{\bf l}=n\tilde{\Phi}_{E}^{0}\;.
\label{fluxoid}
\end{eqnarray}
$\tilde{\Phi}_{E}$ should be the proper definition of a fluxoid that is always quantized in units of $\tilde{\Phi}_{E}^{0}=h/\lambda|{\boldsymbol\mu}|$ in the presence of any planar ${\bf E}$ field. The flux quantum $\tilde{\Phi}_{E}^{0}$ is the same as that defined for generic spintronic devices \cite{Chen13}, since the quantization condition, Eq. (\ref{fluxoid}), has the same relativistic origin as the flux quantization in spintronic devices. The flux ${\boldsymbol\Phi}_{E}$ is a vector defined by the cross product of {\bf E} field and the trajectory
\begin{eqnarray}
{\boldsymbol\Phi}_{E}=\oint{\bf E}\times d{\bf l} .
\label{Phi_E_definition}
\end{eqnarray}
At a given ${\bf E}$ field, the spin wave function, represented by the local angles $\left\{\theta_{i}\right\}$ of the spin orientation, is determined by the requirement of single-valuedness Eq. (\ref{single_valuedness}), and by satisfying the continuity condition Eq. (\ref{continuity_condition}) and minimizing the energy of the whole system. The winding number $n$ and spin current ${\bf J}_{s}$ can then be calculated from $\left\{\theta_{i}\right\}$.

An illustrative example is the vortex \cite{Mostovoy06} induced by ${\bf E}={\bf{\hat r}}\eta/2\pi\epsilon_{0}r$ of a line charge of density $\eta$  through the center of the Corbino ring. At radius $r\gg a$, $n$ is determined by minimizing the energy per bond on the ring with circular geometry around the line,
\begin{eqnarray}
-\cos\left(\delta\theta_{n}-\theta_{E}\right)\approx -1+\frac{a^{2}}{2r^{2}}\left(n-\frac{\eta}{\epsilon_{0}\tilde{\Phi}_{E}^{0}}\right)^{2}\;,
\label{En_symmetric_Corbino}
\end{eqnarray}
where the quantized $\delta\theta_{n}=  a {\boldsymbol{\hat\theta}}\cdot {\boldsymbol\nabla}\theta =na/r$ follows from Eq. (\ref{single_valuedness}), and $\theta_{E}=\lambda\left({\boldsymbol\mu}\times{\bf E}\right)\cdot {\boldsymbol{\hat\theta}}a/\hbar$ in this case. So $n$ is independent of $r$. The spin current is 
\begin{eqnarray}
{\bf J}_{s}={\boldsymbol{\hat \theta}}\frac{\tilde{J}S^{2}}{a\hbar r}\left(n-\frac{\eta}{\epsilon_{0}\tilde{\Phi}_{E}^{0}}\right)\;.
\end{eqnarray}
Therefore the vortex has a spin current that circulates the line charge and decreases with distance. The winding number and the corresponding spin texture can be tuned by the line charge, and should be measurable by polarization-sensitive probes such as optical Kerr effect \cite{Pechan05}, Lorentz transmission electron microscopy \cite{Graef00}, or magnetic transmission soft X-ray microscopy \cite{Fischer08}. Compared to several existing proposals to generate a spin current on a ring \cite{Schutz03,Wu05}, the spin current created in this way is a ground state property that does not require the excitation of magnons.

In this context we may also ask
whether the spins would tilt out of the plane in order to lower the energy when $\delta\theta_{n}\neq\theta_{E}$. Consider the state $|n^{b}\rangle$ where the spin at site $i$ tilts out of plane with a small angle $\gamma_{i}$. The energy on the bond $\left\{i,i+\delta\right\}$ is 
\begin{eqnarray}
E_{i,i+\delta}^{b}=\tilde{J}S^{2}\cos\left(\delta\theta_{n}-\theta_{E}\right)+\Delta E_{i,i+\delta}^{b}
\end{eqnarray}
where
\begin{eqnarray}
\Delta E_{i,i+\delta}^{b}\approx-\frac{\tilde{J}}{2}S^{2}\left(\gamma_{i}^{2}+\gamma_{i+\delta}^{2}\right)\cos\left(\delta\theta_{n}-\theta_{E}\right)
+JS^{2}\gamma_{i}\gamma_{i+\delta}
\nonumber \\
\approx\frac{|J|}{2}S^{2}\left\{\left(\gamma_{i}-\gamma_{i+\delta}\right)^{2}
+\left(\gamma_{i}^{2}+\gamma_{i+\delta}^{2}\right)\delta\theta_{n}\left(\theta_{E}-\frac{\delta\theta_{n}}{2}\right)\right\}\;.
\nonumber \\
\end{eqnarray}
Since $\Delta E_{i,i+\delta}^{b}>0$ for $\delta\theta_{n}\neq 0$ and $\theta_{E}>\delta\theta_{n}/2$, tilting is energetically unfavorable for all $n\neq 0$, and the spin order remain coplanar.


\subsection{Little-Parks effect}

The ground state energy described by Eq. (\ref{En_symmetric_Corbino}) is similar to the situation of the superconducting condensate in the Little-Parks experiment \cite{Little62}. Indeed, minimizing Eq. (\ref{GL_DM}) at { radius $r$ on the Corbino ring yields}
\begin{eqnarray}
S^{2}=S_{\infty}^{2}\left[1-\frac{\left(\delta\theta_{n}-\theta_{E}\right)^{2}}{2\left(1-\tau/\tau_{c}\right)}\right]
\end{eqnarray}
where $S_{\infty}^{2}=|\tilde{J}|v\left(1-\tau/\tau_{c}\right)/\beta$ is the spin magnetization in the absence of the radial ${\bf E}$ field. Thus the magnetization shrinks, if the electric flux is not equal to an integer multiple of the flux quantum $ \tilde{\Phi}_{E}^{0} $, and oscillates periodically with $\theta_{E}$, or equivalently with ${\bf E}$ in the small field regime. This is precisely the analog of Little-Parks effect where the magnitude of SC order parameter oscillates periodically with magnetic flux, and should be observable as an oscillation of $\tau_{c}$ as function of electric field.

\begin{figure}[ht]
\begin{center}
\includegraphics[clip=true,width=0.99\columnwidth]{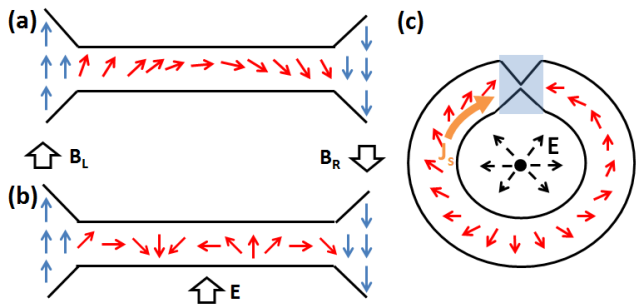}
\caption{ (color online) (a) { A spin Josephson junction consists of a constriction connecting two planar bulk ferromagnets that have misaligned  magnetizations (blue arrows). The weakly coupled spins in the constriction (red arrows) gradually rotate from the configuration on the left to that on the right. The resulting phase gradient creates a spin current. (b) Applying a coplanar ${\bf E}$ in the constriction yields additional rotations on the spin configuration (here $n=1$). (c) Proposed rf SQUID-like device that uses the ${\bf E}$ field from a line charge to control the spin supercurrent $J_{s}$ (orange arrow), where the Josephson junction is a constriction (coated with metal to shield the constriction).}  Arrows (red) indicate the spiral configuration caused by the ${\bf E}$ field.  }
\label{fig:Josephson_open_trajectory}
\end{center}
\end{figure}

\subsection{Spin Josephson junction}

Josephson effect is another example that manifests superfluidity. It has been proposed that magnetic insulators producing triplon condensates can display a spin Josephson effect \cite{Schilling12}. Here we demonstrate that a spin Josephson effect can also be realized in our system by joining two planar ferromagnets with misaligned spin order through a constriction. This effect originates solely from the angular difference between two magnetic orders. Consider the device shown in Fig. \ref{fig:Josephson_open_trajectory}(a), where the constriction is shorter than $\xi$ and has a coupling much smaller than the bulk coupling, $J_{int}\ll J$. The misalignment between ${\bf S}_{L}$ and ${\bf S}_{R}$ (blue arrows in Fig.~\ref{fig:Josephson_open_trajectory}(a)) can be achieved by bulk anisotropy or an external magnetic field. The Josephson spin current can be calculated from the GL free energy, Eq. (\ref{GL_spin_Josephson}), in the { constriction region \cite{Chen13}.} Applying an ${\bf E}$ field further changes the phase gradient and the Josephson spin current (Fig. \ref{fig:Josephson_open_trajectory}(b)). To see this, first we define ${\bf k}_{0}=\lambda{\boldsymbol \mu}\times{\bf E}/\hbar$, whose component $k_{0x}$ points along the junction. The dimensionless quantity $g(x)=\psi(x)/\psi_{\infty}=\psi_{x}/(-\alpha/\beta)$ in the constriction $0<x<L$ where the {\bf E} field is applied satisfies the equation, \cite{Tinkham96}
\begin{eqnarray}
\left(\partial_{x}-ik_{0x}\right)^{2}g=0
\end{eqnarray}
The solution is 
\begin{eqnarray}
g(x)=(1-\frac{x}{L})e^{ik_{0x}x}+\frac{x}{L}e^{-ik_{0x}(L-x)+i\theta_{LR}}
\label{gx_interface}
\end{eqnarray}
such that it satisfies the boundary condition $g(0)=1$ and $g(L)=e^{i\theta_{LR}}$, where $\theta_{LR}$ is the phase (angular) difference of the junction. The current from Eq.~(\ref{GL_spin_Josephson}) yields
\begin{eqnarray}
J_{s}=J_{s}^{0}\sin\left(\theta_{LR}-\varphi_{\rm AC}\right)
\label{Js_FET}
\end{eqnarray} 
where the AC phase is given by
\begin{eqnarray}
\varphi_{\rm AC}=\left(\frac{\lambda|{\boldsymbol\mu}|}{\hbar}\right){\boldsymbol{\hat\mu}}\cdot\int {\bf E}\times d{\bf l}=\frac{2\pi{\boldsymbol{\hat\mu}}\cdot{\boldsymbol\Phi}_{E}}{\tilde{\Phi}_{E}^{0}}\;.
\end{eqnarray}
The periodicity of Eq. (\ref{Js_FET}) again implies that ${\boldsymbol\Phi}_{E}$ is quantized in units of $\tilde{\Phi}_{E}^{0}$. Eq. (\ref{Js_FET}) indicates a $\varphi$-junction whose current-phase relation is controllable by a gate voltage.
As shown in Fig. \ref{fig:Josephson_open_trajectory}(b), $\varphi_{\rm AC}$ also changes the spin configuration like in a domain wall, which should be visible by spin polarization-sensitive probes \cite{Pechan05,Graef00,Fischer08} as a fringe pattern. 
Note that for practical purpose we can change  $\oint\rightarrow\int_{0}^{L}$ in the definition of ${\boldsymbol\Phi}_{E}$ described by Eq. (\ref{Phi_E_definition}) corresponding to an open trajectory. This reflects the fact that, under controlled conditions, quantization of ${\boldsymbol\Phi}_{E}$ can make sense in open trajectory devices too, because the applied field does not display a gauge freedom \cite{Chen13}.

\subsection{Quantum interference in a SQUID geometry}

Interference of the condensate can also take place in the rf SQUID-like geometry proposed in Fig. \ref{fig:Josephson_open_trajectory}(c). The ${\bf E}$ field from the line charge causes a spiral configuration (red arrows in Fig. \ref{fig:Josephson_open_trajectory}(c)), and the weak link can be implemented by a constriction. The phase gained along the path forces the spin supercurrent to oscillate upon increasing ${\boldsymbol\Phi}_{E}$
\begin{eqnarray}
J_{s}=J_{s}^{0}\sin\varphi_{\rm AC}\;.
\end{eqnarray}
For an FM spiral, changing of the spiral wave length through the variation of the ${\bf E}$ field should also be observable by polarization-sensitive probes\cite{Pechan05,Graef00,Fischer08}.


\subsection{Possible experimental realizations}

Several multiferroic compounds exhibiting coplanar spiral order are known, most notably the perovskites $R$MnO$_{3}$\cite{Kimura05,Yamasaki08,Choi10} and $R$FeO$_{3}$ \cite{Tokunaga08,Tokunaga09} (R=Dy, Gd, etc.). The existence of long range spiral order seems to imply that dipole-dipole interaction and lattice anisotropy can be ignored, as assumed in our calculation. Experimental control of spin helicity by ${\bf E}$ field has also been demonstrated \cite{Yamasaki07,Seki08}. We anticipate that the proposed effects can be realized in some of these materials that exhibit DM interactions. The intrinsic spiral typically has wave length $\sim 10a$ along a certain crystalline direction and, thus, would not contribute a phase winding in the Corbino ring or the rf SQUIF-like geometry. While this may not be a conceptual problem at first sight, it for practical purpose desirable to remove the intrinsic spiral by a uniform ${\bf E}$ field if possible, in order observe the proposed effects induced by the line charge.
A quantitative estimate of $w$ is difficult to date.  Nevertheless, on a ring of $\mu$m size, creating $n\sim{\cal O}(1)$ by ${\bf E}$ field is likely within an experimentally accessible range, and the resulting spiral wave length is within the resolution of polarization-sensitive probes.

\section{Conclusions}

By representing a multiferroic with coplanar spin order in the language of a superfluid, we predict several phenomena induced by the phase gradient of the condensate wave function (texture of the spin order) and the coupling to external ${\bf E}$ field. In principle, phenomena caused by $U(1)$ gauge field in superconductors can find their analog in such multiferroics, as long as gauge freedom is not required. 
For this reason there is no analogue of Meissner screening of an external electric field which relies on gauge freedom. In suitable geometries, the ${\bf E}$ field creates topologically distinct spin textures that can be detected by polarization-sensitive probes. 

In this context one may also think about potential applications. Considering, for instance, superconducting qubits. Coplanar magnetic multiferroics may also achieve a similar design, despite some obvious difficulties, by extending the concept of rf SQUID illustrated in Fig.~\ref{fig:Josephson_open_trajectory}(c). The advantage of using multiferroics is that materials with $ \tau_c $ at or above room temperature are available, hence the possibility of room temperature operation. The problem concerning possible applications in quantum computation, as well as the possibility of converting the spin supercurrent inside a multiferroic to a spin current measurable by transport experiments, will be discussed in forthcoming studies.

We thank P. Horsch, D. Manske, M. Kl\"{a}ui, O. P. Sushkov, Y. Tserkovnyak, H. Nakamura, and S. Ishiwata for stimulating discussions. This work was supported by the visitor program of the Pauli Centre for Theoretical studies of ETH Zurich.

\end{document}